\documentstyle[12pt]{article}
\newcommand{\ba}{\begin{array}}
\newcommand{\ea}{\end{array}}

\newcommand{\be}{\begin{equation}}
\newcommand{\ee}{\end{equation}}
\newcommand{\bea}{\begin{eqnarray}}
\newcommand{\eea}{\end{eqnarray}}
\newcommand{\beal}{\setcounter{letter}{1} \begin{eqnarray}}
\newcommand{\eeal}{\addtocounter{equation}{1} \end{eqnarray}}
\newcommand{\none}{\nonumber \\}
\newcommand{\req}[1]{Eq.(\ref{#1})}

\newcommand{\larrow}{\,\,\,\,\hbox to 30pt{\rightarrowfill}
\,\,\,\,}
\newcommand{\slarrow}{\,\,\,\hbox to 20pt{\rightarrowfill}
\,\,\,}

\newcommand{\half}{{1\over2}}

\begin{document}
\mbox{}\\[1cm]
\begin{center}
{\bf CLASSICAL AND QUANTUM MECHANICS OF BLACK\\ HOLES
IN GENERIC 2-D DILATON GRAVITY\footnote{Talk given by
G. Kunstatter at the Conference on Heat Kernels and Quantum Gravity, Winnipeg,
August, 1994}}
\end{center}
\vspace{0.5 cm}
\begin{center}
{\sl by\\
}
\vspace*{0.50cm}
{\bf
J. Gegenberg $\dagger$
G. Kunstatter $\sharp$ and D. Louis-Martinez $\sharp$ $\flat$
\\}
\vspace*{0.50cm}
{\sl
$\dagger$ Dept. of Mathematics and Statistics,
University of New Brunswick, \\
Fredericton, New Brunswick, Canada  E3B 5A3\\
{[e-mail: lenin@math.unb.ca]}\\ [5pt]
}
{\sl
$\sharp$ Dept. of Physics and Winnipeg Institute of
Theoretical Physics, University of Winnipeg\\
Winnipeg, Manitoba, Canada R3B 2E9\\
{[e-mail: gabor@theory.uwinnipeg.ca]}\\[5pt]
}
{\sl
$\flat$ Dept. of Physics, University of Manitoba\\
Winnipeg, Manitoba, Canada R3T 2N2\\
{[e-mail: martinez@uwpg02.uwinnipeg.ca ]}
}
\end{center}
\bigskip\noindent
{\large
ABSTRACT
}
\par
A unified description is presented of the physical observables and
thermodynamic variables
associated with black hole solutions in generic 2-D dilaton gravity. The Dirac
quantization of these theories is reviewed and an intriguing relationship
between the entropy of the black hole and the WKB phase of the corresponding
physical wave functionals is revealed.

\newpage
\renewcommand{\baselinestretch}{1.5}
\section{Introduction}\medskip
\par
One of the most difficult and important goals of theoretical
physics is to
construct a mathematically and conceptually consistent theory of
quantum gravity. The inadequacy of the  usual semi-classical
approximation, in which only matter fields are quantized, has
recently been emphasized in the context
of the so-called information loss paradox associated with the
Hawking radiation of black holes\cite{info}. There have been many attempts in
recent years to avoid the pitfalls of the usual semi-classical approximation by
 considering models of quantum gravity in two spacetime dimensions. In the
absence of matter, these models are exactly solvable both classically and
quantum mechanically. The hope is that, in addition to providing  useful
conceptual and theoretical insights, in some cases they  provide reasonable
approximations to the quantum mechanical behaviour of physical
black holes. One model in particular, (spherically symmetric gravity or SSG) is
 obtained by truncating all non-spherically symmetric modes in ordinary
Einstein gravity in 3+1 dimensions.\cite{ssg}. Another model of current
interest is string inspired 2-d dilaton gravity(SIG) which was recently used to
investigate perturbative effects of backreaction on the
end-point of gravitational collapse\cite{cghs}. Finally,  Jackiw-Teitelboim
gravity (JT)\cite{JT}, which was the first 2-D dilaton theory, can  be obtained
by imposing axial symmetry in
2+1 Einstein gravity. The latter has gained new importance in the present
context due to the recent discovery\cite{BTZ} of black hole  solutions in the
2+1 dimensional theory. These solutions have been used in an attempt to provide
a statistical origin for black hole entropy\cite{carlip}.
\par
The models mentioned above are all special
cases of generic (1+1) dimensional dilaton gravity theories\cite{banks} in
which a scalar field is non-minimally coupled to the spacetime curvature. As
will be shown below, it is possible to provide a complete, unified treatment of
all the models by adapting a suitable parametrization for the fields. In
particular, we will be able to provide explicit representations for the
classical observables for black holes (if they exist) in all models including
the phase space variables, the surface gravity and the entropy. Moreover, it is
possible to solve the constraints exactly in these models and write down
physical quantum wavefunctions\cite{domingo1} which can be interpreted as  a
WKB approximation to the stationary states describing quantum black holes. One
of the more intriguing results that emerges from this analysis is the
relationship between the imaginary part of the WKB phase and the classical
thermodynamic entropy\cite{kkm}.
\par
The paper is organized as follows. In Section 2 we present the action
for generic 2-D dilaton gravity and discuss the space of the
solutions. It turns out that the theories obey a generalized Birkhoff
theorem\cite{domingo2}: all solutions have a timelike Killing vector,
and  can be parametrized by the value of a single coordinate
invariant, conserved quantity(i.e the energy). Section 3 describes
the classical thermodynamics for these models and shows how to
calculate the surface gravity and entropy in arbitrary coordinates
for the generic theory. Section 4 summarizes the canonical analysis
for these theories, showing that the reduced phase space is two
dimensional despite the fact that there exists only a one parameter
family (up to space time diffeomorphism) of inequivalent solutions.
The physical interpretation of these two phase space variables is
also presented: as expected one is the energy while its conjugate is
related to the Killing time separation between slices at spatial
infinity. This interpretation was recently given for
    SSG by Kuchar\cite{kuchar}.  Section 5 reviews the Dirac
quantization of the models and shows the intriguing, generic
relationship between the black hole entropy and the WKB phase.
Section 6 closes with conclusions and prospects for future research.

\section{Action and Space of Solutions}\medskip
The most general action functional depending on the metric tensor $g_{\mu\nu}$
and a scalar field $\phi$ in two spacetime dimensions, such that it contains at
most second derivatives of the fields can be written\cite{banks}:
\be
S[g,\phi] ={1\over 2G} \int d^2x \sqrt{-g} \left(\half  g^{\alpha\beta}
\partial_\alpha\phi\partial_\beta\phi + {1\over l^2} \tilde{V}(\phi) +
D(\phi) R \right)\,\, .
\label{eq: action 1}
\ee
In the above, $G$ is the (dimensionless) 2-d gravitational constant and
$D(\phi)$ and $\tilde{V}(\phi)$ are arbitrary functions of the scalar field.
This action is a generalization of what occurs naturally if one restricts to
only spherically symmetric modes in 3+1 Einstein gravity. To see this consider
the spherically symmetric metric
\be
ds^2=g_{\mu\nu}(x)dx^\mu dx^\nu + e^{-2\phi(x)}d\Omega^2,
\ee
where the $x^\mu$ are coordinates on a two dimensional spacetime $M_2$
with metric $g_{\mu\nu}(x)$ and $d\Omega^2$ is the line element of the 2-sphere
with area $4\pi$.  The spherically symmetric vacuum Einstein solutions
are the stationary points of the dimensionally reduced action functional:
\be
\tilde I[g,\phi]={1\over 2G}\int_{M_2} d^2x \sqrt{-g}
e^{-2\phi}\left(R(g)+2g^{\mu\nu}
\partial_\mu\phi\partial_\nu\phi +2 e^{2\phi}\right).
\ee
which is of the same form as \req{eq: action 1} above.  As first discussed in
\cite{banks} and shown explicitly in
\cite{domingo1}, one can
eliminate the kinetic
term for the scalar by reparametrizing the  fields:
\bea
 g_{\mu\nu} &\to& h_{\mu\nu}=\Omega^2(\phi) g_{\mu\nu}
\label{eq: gbar defn},\\
\phi &\to& \tau=D(\phi),
\eea
with
$\Omega^2=\exp\half\int{d\phi\over(dD/d\phi)}$. This leads to an action
functional of the form:
\be
I[h,\tau]={1\over2G}\int_{M^2}d^2x \sqrt{-h}\left(\tau R(h)+{1\over
l^2}{V(\tau)}
\right).\label{eq: dilaton action}
\ee
where $V(\tau)$ is an arbitrary function of the scalar field
$\tau$. For spherically symmetric gravity, $V(\tau)= 1/\sqrt{2\tau}$,
while  $V(\tau)= \tau$ for JT,and  $V(\tau)=1$ for SIG.
\par
The equations of motion take the simple form:
\be
R=-{1\over l^2} {dV\over d\tau} \label{eq:eofm1}
,\ee
and
\be
\nabla_\mu\nabla_\nu\tau-{1\over 2l^2} g_{\mu\nu}V=0
\label{eq:eofm2}.
\ee
\par
As shown in \cite{domingo2} the theory obeys a generalized Birkhoff's theorem
which states that up to spacetime diffeomorphisms there is a one parameter
family of solutions for each choice of potential. The parameter is a
coordinate-invariant constant of integration which can be expressed
in covariant form as follows:
\be
E = -{1\over 2l}((\nabla\tau)^2 l^2+ J(\tau) )\,\, .
\label{eq: covariant 2lE}
\ee
It will be shown in Section 4 that $E$ can be interpreted as the energy of the
solution, as expected.
\par
 The solutions take a particularly simple form in what we call the
Schwarzschild gauge (since the relationship to SSG is manifest). Choosing
\be
\tau=x/l, g_{tx}=0,
\ee
the most general solution is:
\be
ds^2=-({J(x/l) }-2lE)dt^2+({J(x/l)}-2lE)^{-1}dx^2 \,\,\, ,
\label{eq: generic solution}
\ee
where $J'(\tau)= V(\tau)$
\par
The solutions as written above clearly have a Killing vector, which can also be
written in covariant form:
\be
k^\mu=l\eta^{\mu\nu}\tau,{}_\nu \label{eq:killing} \,\, .
\label{eq: killing vector}
\ee
In the above
$\eta^{\mu\nu}=-\eta^{\nu\mu}={1\over\sqrt{-g}}\epsilon^{\mu\nu}$
is the antisymmetric {\it tensor}. The constant $l$ is required to
make the Killing vector components dimensionless.
It can easily be verified that \req{eq:eofm2} implies that $k^\mu$
satisfies the Killing equation $\nabla_{(\mu}k_{\nu)}=0 $ on shell.
Moreover, it is clear that
$\tau_{,\mu}k^\mu=0$ identically, so that the scalar field is also
invariant along the Killing directions. Note that
\be
\mid k\mid^2=-l^2\mid\nabla\tau\mid^2\none
            = 2lE-J(\tau)  \,\, .
\label{eq:magkilling}
\ee
\par
Both the form of the solution in Schwarzschild gauge, and the magnitude of the
Killing vector suggest that
the dilaton
theories
admit black holes providing that there exists at least one curve in spacetime
given by
$\tau(x,t)=\tau_0=constant$, such that $J(\tau_0)=2lE$. In addition,
$J(\tau)$ must be monotonic (in $\tau$) in a neighbourhood of
$\tau_0$. These conditions are satisified for SSG, JT and SIG, as well as
for a large class of additional theories\cite{portuguese}.
For example in SSG, for which $V(\tau)=1/\sqrt{2\tau}$,  the static solution
for the
metric in our parametrization  is related
to the usual Schwarzschild solution by the conformal
reparametrization $ds^2 = \sqrt{2\tau}ds^2_{schwarz}$. In terms of
the coordinate
$r=l\sqrt{2\tau}$, the metric \req{eq: generic solution} takes the
form:
\be
h_{\mu\nu}dx^\mu dx^\nu={r\over l}\left\{-(1-2m/r)dt^2+(1-2m/r)^{-
1}\right\}dr^2,
\ee
where the mass $m=l^2 E$. Finally, $(k^\mu)=(1,0)$ and
$|k|^2 = (2m-r)/l$.
\par
 For JT,  we define $\tau= x/l$, so that the solution takes the form:
\be
h_{\mu\nu}dx^\mu dx^\nu=\left\{-({x^2\over l^2}-2lE)dt^2+({x^2\over
l^2}-2lE)^{-
1}\right\}dx^2,
\ee
The black hole in this case is the 2-D projection of the BTZ solution
\cite{BTZ} in 2+1 gravity: it is not asymptotically flat, because the metric
describes a spacetime that is everywhere locally de-Sitter.
\par
Finally, for the string inspired theory, with $\tau=x/l$ as above:
\be
h_{\mu\nu}dx^\mu dx^\nu=\left\{-({x\over l}-2lE)dt^2+({x\over l}-2lE)^{-
1}\right\}dr^2,
\label{eq: flat sig metric}
\ee
What is remarkable is that the metric in the above solution is completely flat
in the parametrization we have chosen. This is a consequence of the fact that
in 2 spacetime dimensions all metrics are conformally flat:  the
reparametrization \req{eq: gbar defn} has in this case transformed the physical
metric to the flat metric in \req{eq: flat sig metric}. As  shown below,
however, the information about the event horizon and associated thermal
properties is still encoded in the solution, as long as one considers both the
scalar field and the metric.
\par
 Note that at this stage
we are discussing only local properties of the solutions. To establish
rigorously the existence of an event horizon one needs to study global
properties. In the three theories of immediate interest, these global
properties do support the existence of event horizons. With a slight abuse of
notation we will generically denote   the curve defined by  $J(\tau_0)=2lE$ as
the event horizon, since this is the curve along which the Killing vector is
null.
\bigskip
\section{Thermodynamical Properties}
\medskip
Given the possible existence of black holes and event horizons, it is
straightforward to use the explicit expression for the Killing vector to
calculate the surface gravity and entropy of a generic 2-D
black hole. The surface gravity $\kappa$
is determined by the following expression,
evaluated at the event horizon\cite{wald2}:
\be
\kappa^2=-{1\over2}\nabla^\mu k^\nu\nabla_\mu k_\nu.
\ee
Using \req{eq: killing vector} for $k^\mu$ and the field equations
\req{eq:eofm2}one obtains:
\be
\kappa =  {1\over2 l}V(\tau_0) \,\, ,
\label{eq: kappa}
\ee
where $V(\tau_0)$ is the potential evaluated at $\tau=\tau_0$ (i.e.
on the event horizon).
The sign in \req{eq: kappa} was chosen to yield a positive surface
gravity for positive energy. Note that $\tau_0$ is given implicitly
as a function of the energy $q$ by requiring $|k|^2=0$ in
\req{eq:cov q}.
\par
The Hawking temperature for the generic black hole solution can
 be calculated heuristically by defining the Euclidean time $t_E = it$ in
\req{eq: generic solution} and then finding the periodicity
condition on $t_E$ that makes the solution everywhere regular. This
is done by defining the coordinate $R^2:= a (J(\tau)-2lE)$  and
choosing the constant $a$ so that the spatial part of the metric
goes to $dR^2$ at the event horizon $\tau_0$. A straightforward
calculation gives $a=|2l/V(\tau_0)|$, so that the Hawking
temperature,
which is the inverse of the period of $t_E$, is:
\be
T_H = {1\over 2\pi} {V(\tau_0)\over 2l} = {\kappa\over 2\pi} \,\,
,
\ee
as expected. Note that this calculation does not depend on the
details of the model: it merely requires the existence of a horizon
at which $J(\tau_0) = 2lE$
\par
The entropy, $S$, can now  be determined by inspection of
\req{eq:magkilling}. In particular, if one varies the solution, but stays
on the event horizon  (i.e. at $\tau=\tau_0$),  one finds the variation of the
energy to be:
\be
\delta E ={1\over 2l G}\delta J(\tau_0)= {1\over 2l G} V(\tau_0) \delta
\tau_0 \,\, .
\label{eq: delta q}
\ee
Identifying the Hawking temperature and surface gravity derived
above, we find that the first law of thermodynamics
$\delta E = T\delta S$ will be satisfied providing we identify
the entropy to be
\be
S = {2\pi\over G} \tau_0 \,\, .
\label{eq: entropy}
\ee
This expression agrees with  Wald's\cite{wald} more general local geometric
formulation for the
entropy of a black hole. For SIG $V(\tau)= 1/\sqrt{2\tau}$, and for the
solution given above  $\tau_0= 2 m^2/l^2_p$. Thus \req{eq: entropy} gives the
correct entropy for a spherically symmetric black hole. It also  agrees
with results for black holes in SIG\cite{frolov}\cite{iyer} and for the
dimensionally reduced BTZ black hole\cite{BTZ} obtained as a solution to JT.
\par\bigskip\noindent
\section{Hamiltonian Analysis}
\medskip
We now review the Hamiltonian analysis of the general
1+1--dimensional theory\cite{domingo1}.
Spacetime is split into
a product:  $M_2 \simeq
\Sigma\times
R$ and the metric $h_{\mu\nu}$ is given an ADM-like
parameterization:\cite{torre}.\be
ds^2=e^\alpha\left[-\left(M^2+N^2\right)dt^2+\left(dx+Mdt\right)
^2\right]
.\label{eq:adm}
\ee
where $\alpha$, $M$
and
$N$ are functions on spacetime $M_2$.
We define the quantity $\sigma$ by $\sigma^2:= M^2+N^2$.  Also, in
the following, we denote by the overdot and prime, respectively,
derivatives with respect to the time coordinate $t$ and spatial
coordinate $x$.
\par
The canonical momenta for the fields
$\{\alpha,\tau\}$ are respectively:
\bea
\Pi_\alpha&=&{1\over2G\sigma}\left(M\tau'-\dot\tau\right),
\label{eq: pi_alpha}
\\
\Pi_\tau&=&{1\over2G\sigma}\left(-\dot\alpha+M\alpha'+2M'
\right),
\label{eq: pi_tau}
\eea
The vanishing
of the momenta canonically conjugate to $M$ and $\sigma$
yield the primary constraints for the system. By following the
standard
Dirac prescription\cite{dirac}, it can be shown\cite{domingo2} that the
canonical Hamiltonian is, up to a surface term, a linear combination of
first class constraints:
\be
H_0 = {1\over 2G}\int dx \left[\left({\dot{\tau}\over \tau'}\right) {\cal F}
   - \left( {\sigma e^\alpha \over l \tau'}\right){\tilde{\cal G}}\right]dx\,\,
  {}.
\ee
where  $\dot\tau$ is to be considered as a function of the phase
space coordinates and lapse and shift functions, given implicitly by
\req{eq: pi_alpha}. The constraints ${\cal F}$ and $\tilde{\cal G}$ generate
spatial diffeomorphism and translations along the Killing vector\cite{kkm},
respectively.
They are:
\be {\cal F}:=\alpha' \Pi_\alpha+\tau' \Pi_\tau-2\Pi_\alpha'
\label{eq: gauss}
 \ee
\be
\tilde{\cal G}:=(q[\alpha,\tau,\Pi_\alpha,\Pi_\tau])'
\approx 0 \,\, ,
\label{eq: def'n q'}
\ee
where we have defined the variable $q[\alpha,\tau]$ as
\be
q[\alpha,\tau]:={l\over 2}\left[e^{-\alpha}\left((2G\Pi_\alpha)^2-
     (\tau')^2\right)+ l^{-2}J(\tau)\right].\label{eq:qdef}
\ee
The expression on the right-hand side above is nominally an
implicit function
of the spatial
coordinate, but is constant on the constraint surface. Moreover,
it is straightforward to show
that $q$ commutes with both
constraints ${\cal F}, \tilde{\cal G}$. Thus, the constant mode of $q$ is a
physical
observable in the Dirac sense.
\par
In order for the variation of the above Hamiltonian to yield the correct
equations of motion it is necessary to add the following surface term:
\be
H_{ADM} = \int dx \left(\left( {\sigma e^\alpha \over l
\tau'}\right){q\over G}\right)' \,\, .
\ee
It is easy to verify that for solutions of the form \req{eq:
generic solution},
${\sigma e^\alpha / l \tau'}=1$. Hence, $H_{ADM} = q/G$ is the ADM
energy,
as expected.
\par
In terms of the canonical momenta the magnitude of the Killing
vector can be written as:
\be \mid
k\mid^2=l^2e^{-\alpha}\left[(2G\Pi_\alpha)^2-(\tau')^2\right] \,\,
{}.
\label{eq:kill/mom} \ee
Thus the observable $q$ is:
\bea
q&=&{1\over 2l}\left(\mid k\mid^2+J(\tau)\right)\none
  &=& E \,\, .
\label{eq:cov q}
\eea
This proves that the parameter $E$ defined in Section 2 is one of the physical
phase space observables in the theory, and that it is the  energy of the
solution as anticipated in Section 2.
\par
The momentum conjugate to $q$, is found by inspection to
be\cite{domingo1}:
\be
p:=-\int_\Sigma dx {2\Pi_\alpha e^\alpha\over
(2G\Pi_\alpha)^2-(\tau')^2} \,\, .
\label{eq:pdef}
\ee
It can easily be verified that the Poisson algebra for the fields
and the momenta leads directly to $\{q,p\} = 1$. Moreover, under  a
general gauge transformation the change in $p$ is the integral of a total
divergence\cite{kkm}, which vanishes only if the test functions go to zero
sufficiently rapidly at
infinity. The value of $p$ therefore depends on the global properties of the
spacetime slicing. This is consistent with the generalized Birkhoff
theorem\cite{domingo1} which states that there is only one
independent diffeomorphism invariant parameter
characterizing the space of solutions.
\par
The observable $p$ can be written in covariant form\cite{kkm}:
\bea
p&=&-2\int_\Sigma dx^\mu {k_\mu\over\mid
k\mid^2}.\label{eq: cov p}
\eea
This form of the expression shows that in static slicings $p=0$. Moreover, the
integrand of  $p$ has a pole at the location of any event horizon in
the model. Thus, analytic continuation is in general required to
make the expression well defined, and may introduce an imaginary
part to the observable $p$. For example, in spherically symmetric
gravity, one can show that in Kruskal coordinates the observable
$p$ integrated along a slice of constant Kruskal time $T$ takes the
simple form:
\bea
p&=&{2m\over G}\int dX\left[{1\over X-T} - {1\over X+T}
\right]\none
 &=&{2m\over G} \left.\ln\left({X-T\over X+T}\right)
\right|^{X_f}_{X_i} \,\,.
\label{eq: p in Kruskal}
\eea
$p$ is therefore precisely the difference in Schwarzschild times
at the
endpoints of the spatial slice. In this case there are simple poles
at $X=\pm T$ (i.e. at $r=2m$), so that for an eternal black hole,
with
suitable analytic continuation, Im${p}= 2\pi m/G$. Although this
potential imaginary piece is irrelevant classically for the
Schwarzschild time, it may have some significance in the quantum
theory in which $p$ is a physical phase space observable.
\par
The
global variable $p$ has a natural physical interpretation in terms of  the time
separation at infinity of
neighbouring spacelike
surfaces  which are asymptotically normal to the Killing vector
field $k^\mu$. Let $U$ be the ``triangular region" of
spacetime bounded by spacelike surfaces $\Sigma_1, \Sigma_2$ and
by a timelike surface $T$ at infinity tangent to $k^\mu$.  By using Gauss' law,
it is possible to show\cite{kkm} that between two such surfaces, the difference
in $p$ is
\be
 p_2-p_1=-\int_T d\theta ,
\ee
where $p_1, p_2$ are the values of $p$ on $\Sigma_1, \Sigma_2$,
respectively, and $\theta$ is a
parametrization of the timelike line $T$ such that the induced
metric $$h_{\theta\theta}:=h_{\mu\nu}{\partial
x^\mu\over\partial\theta}{\partial x^\nu\over\partial\theta}=\mid
k\mid^2$$.
\par
\par\bigskip\noindent
\section{Dirac Quantization}
\medskip
We will now outline the procedure for the quantization of the generic theory in
the functional Schrodinger representation.   Following the Dirac prescription
we first define a Hilbert space in terms of wave functionals
$\psi[\alpha,\tau]$ of
the configuration space coordinates $\alpha(x)$ and $\tau(x)$. The
momentaconjugate to $\alpha$ and $\tau$ are represented as self-adjoint
operators on this Hilbert space. If the Hilbert space measure is:
\be
<\psi|\psi>=\int \prod_xd\alpha(x)d\tau(x)\mu[\alpha,\tau] \psi^*[\alpha, \tau]
  \psi[\alpha,\tau]
\label{eq: measure}
\ee
then
\bea
\hat{\Pi}_\alpha &=&-i\hbar {\delta\over \delta \alpha(x)} + {i\hbar\over 2}
    {\delta  \ln (\mu[\alpha,\tau])\over \delta\alpha }\\
\hat{\Pi}_\tau &=&-i\hbar {\delta\over \delta \tau(x)} + {i\hbar\over 2}
     {\delta \ln (\mu[\alpha,\tau])\over\delta \alpha}
\eea
Physical states are those that are annihilated by the quantum operators
corresponding to the constraints $\cal F$ and $\tilde{\cal G}$. Since the
latter is quadratic in the momentum one has to find a suitable factor ordering
that makes the operator self-adjoint with respect to the chosen functional
measure. In the following we will consider only the lowest order WKB
approximation, which is insensitive to the factor ordering and choice of
measure\footnote{Note that the semi-classical approach taken here is
somewhat different than the one suggested in \cite{domingo1} and \cite{gk}, in
which the constraints were solved classically for the momenta, and then imposed
 exactly as quantum mechanicail constraints on the space of physical states.
The WKB approximation has also recently been discussed in the present context
by Lifschytz {\it et al}\cite{mathur}} . We can therefore set
$\mu[\alpha,\tau]=1$ without loss of generality. The extension of this analysis
to higher order is currently under investigation.\par
We now focus on stationary states of the theory, which we define to be
eigenstates of the energy operator $\hat q$ with constant eigenvalue, $E$:
\be
\hat q \psi[\alpha,\tau] = E \psi[\alpha,\tau]
\label{eq: schrodinger}
\ee
that are annihilated by the diffeomorphism constraint:
\be
\hat {\cal F} \psi[\alpha,\tau] = 0
\label{eq: gauge}
\ee
Note that the eigenvalue equation above is sufficient (although not necessary)
for the state to satisfy the Hamiltonian constraint, since:
\be
\hat{\tilde{\cal G}} \psi[\alpha,\tau] = E' \psi[\alpha,\tau]=0
\ee
Thus, the quantum constraints \req{eq: schrodinger} and \req{eq: gauge} yield
physical states which satisfy:
\be
\hat H \psi[\alpha,\tau] = E \psi[\alpha, \tau]
\ee
where $\hat H$ is the quantized version of $H:= H_0+ H_{ADM}$ given in Section
4. These are therefore the analogue of ``stationary states" in ordinary
non-relativistic quantum mechanics.
\par
To implement the WKB approximation, we assume that the wave functional can be
written:
\be
\psi[\alpha,\tau] = exp {i\over\hbar}[S_0[\alpha,\tau] + \hbar S_1[\alpha,\tau]
   +...]
\ee
To lowest order in $\hbar$ we find that $S_0$ must be invariant under spatial
diffeomorphisms:
\be
\alpha'{\delta S_0\over \delta \alpha} + \tau' {\delta S_0 \over
    \delta \tau} - 2 \left({\delta S_0\over\delta \alpha}\right)'
=0
\ee
and that it obeys the Hamilton Jacobi equation associated with the energy
function $q[\alpha,\tau]$ given in \req{eq:qdef}:
\be
\left[e^{-\alpha}\left((2G \left({\delta S_0 \over \delta\alpha}\right)^2-
     (\tau')^2\right)+ l^{-2}J(\tau)\right] = {2E\over l}.
\label{eq: hamilton-jacobi}
\ee
These two equations can be solved algebraically to yield:
\be
{\delta S_0 \over \delta \alpha} =Q[\alpha,\tau; q]
\label{eq: Q}
\ee
and
\be
{\delta S_0 \over \delta \tau} = {1\over Q[\alpha,\tau]}
  \left( 2\tau'' - \tau'\alpha'-e^\alpha V(\tau) \right)
\ee
where
\be
Q:= \sqrt{ (\tau')^2+ e^\alpha\left({2E\over l}-{J(\tau)\over
l^2}\right)}
\ee

The closure of the constraint algebra guarantees that these functional
differential equations can be integrated. The solution, first given in
\cite{domingo1} is\footnote{In the case of Jackiw-Teitelboim gravity, this
solution has recently been shown\cite{benedict} to be quantum mechanically
equivalent to the wave function obtained by quantizing
the gauge theoretic version of that theory.}:
\be
S_0[\alpha,\tau;q] = \int dx \left[Q+ {\tau'\over2} \ln
\left({\tau' -
    Q\over \tau'+Q}\right)\right] \,\, ,
\label{eq: wave function}
\ee
\par
As usual in the WKB method, the imaginary parts of the phase are related to
quantum mechanical tunnelling into classically forbidden regions, which in this
case occur for $Q[\alpha,\tau] = 0$. This follows from the fact that
$Q=\Pi_\alpha$ on the constraint surface (see \req{eq: Q} above).
It is interesting to note that if we
restrict to classically allowed regions, for which $Q^2\ge 0$ then
the phase $S$ can still acquire an imaginary part from the logarithm when
$(\tau')^2 - Q^2\le 0$. This is precisely the region where the
Killing vector for the solution is spacelike (in a non-singular
coordinate system for which $e^\alpha$ is positive).
Therefore for theories with an event horizon, the logarithm in
\req{eq: wave function} can be analytically continued so that
\be
Im S_0= {i\pi   \tau_0\over 2} = i {S\over 4} \,\, .
\ee
The
imaginary part of the WKB phase is therefore proportional to the
entropy of the black hole.  This is consistent with an earlier
heuristic result obtained for spherically symmetric
gravity\cite{gk} and suggests a relationship between black hole entropy, and
quantum mechanical tunnelling processes.
\section{Conclusions}
\medskip
By choosing a suitable parametrization, we have been
able to present a unified treatment of the most general dilaton
gravity theory, including a complete characterization of the
classical space of solutions, phase space variables and
thermodynamical quantities. We have also outlined the Dirac quantization
of the theory in the functional Schrodinger representation, and calculated the
WKB phase. In the process,  an interesting connection emerged
between the imaginary part of the phase of these
wave functionals and the entropy of the corresponding black
holes.
\par
These results indicate that it may be possible to extract significant
information concerning Hawking radiation, black hole entropy and perhaps even
the endpoint of gravitational collapse from these models.
Before this can be done, however, several issues should be investigated and
clarified. It should be possible to go beyond leading order in the WKB
approximation, defining a suitable Hilbert space measure and verifying the
closure of the constraint algebra\cite{barvinsky}. It is also necessary to
investigate these models with  matter, since Hawking radiation is not strictly
possible in vacuum dilaton gravity due to the absence of radiative modes.
Finally it should be possible to check whether one can provide a statistical
mechanical interpretation of the black hole entropy in these simple models in
terms of quantum states on the horizon, as done recently for the BTZ black
holes by Carlip\cite{carlip}. On the surface, the states in Carlip's analysis
appear to arise due to the presence of non-axially symmetric gauge modes,
negating the possibility of a similar result in the dimensionally reduced case.
However, since black holes in  JT gravity have the same temperature and entropy
as the 2+
   1 case, one might expe
ct to be able to account for this entropy using similar techniques in both
theories. These questions are currently under investigation.
\par\vspace*{10pt}
\noindent
{\large\bf Acknowledgements}
\par
The authors are grateful to
A. Barvinsky, S. Carlip, V. Frolov, D. McManus, P. Sutton, G.A.
Vilkovisky and D.  Vincent for helpful discussions.  This work
was supported in part by the Natural Sciences and Engineering
Research
Council of Canada.  \par
\end{document}